# Speed Plot Determination of the Δ(1232) Parameters

Mohamed E. Kelabi[1], Khaled A. Mazuz[1], Eman O. Farhat[1]

**Abstract**

From the parameterization of the pion-nucleon phase shift of the partial wave $P_{33}$, the pole position and width of the Δ(1232) resonance is determined by using speed plot technique. We found the pole position and the half width of the Δ(1232) resonance at the total center of mass energy $W = (1211.47 - i\,49.94)$ MeV. The results of our calculations are in agreement with the early determinations.

## 1. Introduction

The Δ(1232) can be treated as a resonance in the elastic region. The corresponding partial wave scattering amplitude ($I = J = 3/2$) is given by

$$f = \frac{\sin\delta\, e^{i\delta}}{q}, \qquad l = 1 \qquad (1)$$

where $\delta \equiv \delta(W)$ is the $\pi N$ phase shift, and

$$q^2 = \frac{\left[W^2 - (m_N + m_\pi)^2\right]\left[W^2 - (m_N - m_\pi)^2\right]}{4W^2}$$

is the total center of mass momentum, with total center of mass energy,

$$W^2 = m_N^2 + 2m_N E_\gamma,$$

where $E_\gamma$ is the photon energy in the Laboratory system, $m_N$ and $m_\pi$ are the nucleon and pion rest masses, respectively.

If we rewrite Eq. (1) of the form

$$qf = \frac{\sin\delta}{e^{-i\delta}} = \frac{1}{\cot\delta - i} \qquad (2)$$

and expanding $\cot\delta(W)$ about the resonance value $W = W_r$ ($\delta = \pi/2$), using Taylor expansion, we get

---
[1] Physics Department, Faculty of Science, Al-Fatah University, Tripoli, LIBYA.

$$\cot\delta(W) = \cot\delta(W_r) + (W - W_r)\frac{d}{dW}\cot\delta(W)\bigg|_{W=W_r} + \ldots$$

where $\cot\delta(W_r) = 0$. By defining

$$\frac{2}{\Gamma} \equiv -\frac{d}{dW}\cot\delta(W)\bigg|_{W=W_r}$$

we get

$$\cot(W) \approx (W_r - W)\frac{2}{\Gamma} \qquad (3)$$

providing that $\Gamma \ll W_r$. By substituting Eq. (3) back into Eq (2), we obtain

$$qf = \frac{\Gamma/2}{(W_r - W) - i\Gamma/2} \qquad (4)$$

This is the resonance part of the *T*-matrix element for elastic scattering, and commonly known as Breit-Wigner resonance formula.

At energies $W = W_r \pm \frac{\Gamma_r}{2}$, the scattering amplitude $|qf|^2$ decreases to its half maximum, and $|\Gamma_r|$ gives the width of the resonance peak at its half-height, as shown in Fig. 1.

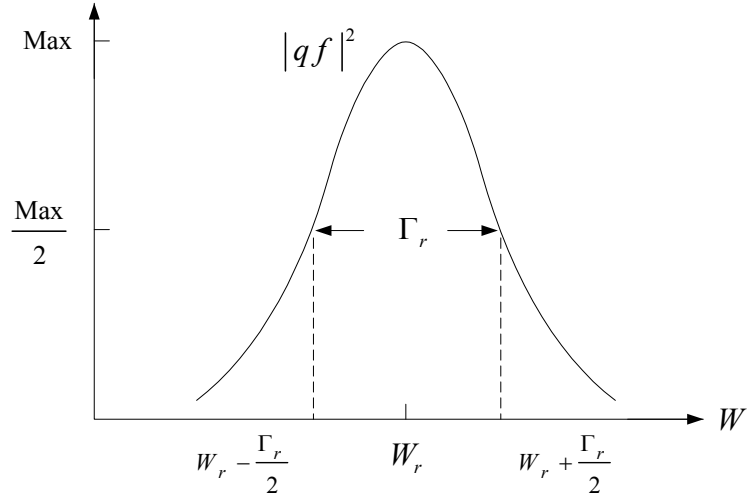

Fig. 1 Graphical representation of Breit-Wigner resonance formula.

In the presence of background, Eq. (4) may take the following form [1]

$$T = \frac{\Gamma/2}{(W_r - W) - i\Gamma/2} + T_B \qquad (5)$$

where the first term represents the resonance amplitude $T_R$, and the background is $T_B$.

## 2. Phase Shift Parameterization

We constrict on accurate parameterization of the phase shift of the form [2]

$$q^3 \cot\delta = a_0 + a_2 q^2 + a_4 q^4 + a_6 q^6 \tag{6}$$

and using an updated version $p\pi^+$ data [3], the following parameters were obtained:

$$\begin{aligned} a_0 &= 5.20252 \\ a_2 &= -0.85689 \\ a_4 &= -0.25211 \\ a_6 &= -0.06119 \end{aligned} \tag{7}$$

in charged pion mass units.

## 3. The Speed Plot method

The method of Speed Plot (*SP*) [4], used by authors to determine the resonance parameters. It is defined by the following relation [5], [6]

$$SP(W) = \left| \frac{dT(W)}{dW} \right| \tag{8}$$

For the single channel case, the background $T_B$ is slowly varying [7] and its derivative can be neglected, then Eq. (5) and Eq. (8) leads to

$$SP(W) = \frac{\Gamma/2}{(W_r - W)^2 + (\Gamma/2)^2}$$

Hence, the maximum of the Speed gives the resonance height [6]

$$SP(W_r) \equiv H = 2/\Gamma_r \tag{9}$$

whereas, the half-maximum value occurs at the Speed [6]

$$SP(W_r \pm \Gamma_r/2) \equiv H/2 = 1/\Gamma_r. \tag{10}$$

## 4. Results

The pole position $W_r$ and width $\Gamma_r$ of the $\Delta(1232)$ resonance are determined by plotting $SP(W)$ verses $W$, and reading off the position at the height of the peak. The resulting Speed plot is shown in Fig. 2, and the resonance parameters obtained from Eq. (9) and Eq. (10) are, respectively:

$$W_r = 1211.47 \text{ MeV} \qquad \Gamma_r = 99.88 \text{ MeV}.$$

These values are comparable with other calculations shown in Table 1.

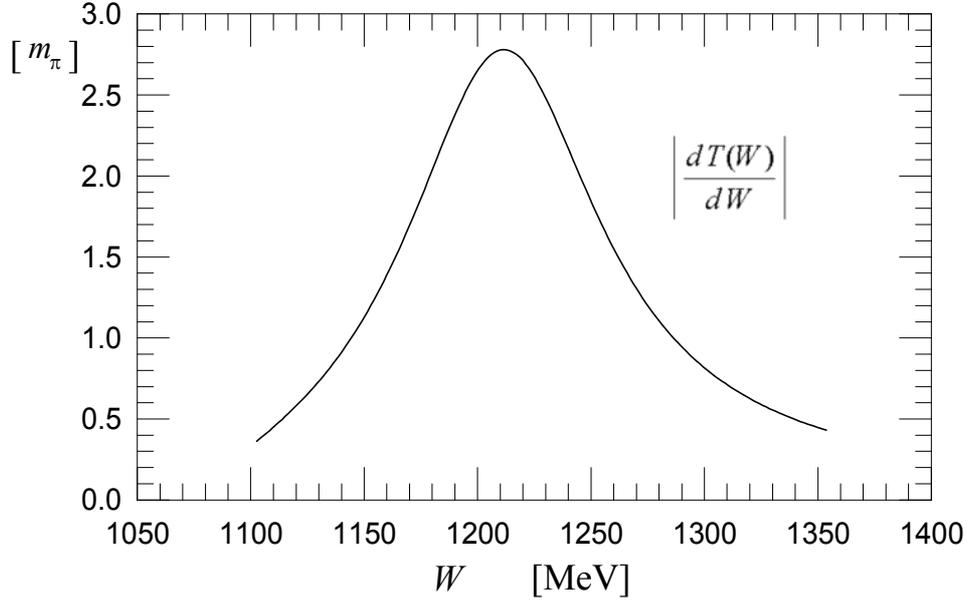

Fig. 2. The speed plot method.

Table 1. Mass and width of the Δ(1232) resonance.

| $W_r$ [MeV] | $\Gamma_r$ [MeV] | Name | Fit |
| --- | --- | --- | --- |
| 1211.47 | 99.88 | Present work | 2010 |
| 1210.8 | 99 | Höhler [8] | 2001 |
| 1211±1 to 1212±1 | 102 ±2 to 99±2 | Hanstein [7] | 1996 |
| 1206.9±0.9 to 1210.5±1.8 | 111.2±2.0 to 116.6±2.2 | Miroshnichenko [9] | 1979 |
| 1208.0±2.0 | 106±4 | Campbell [10] | 1976 |

**Conclusion**

The results obtained of the Δ(1232) resonance from $p\pi^+$ data are in agreement with the early calculations. Our simple parameterization of the phase shift gives accurate fits. However, the accuracy of determining the pole position and width of the Δ(1232) resonance may depend on the data set used.

In the future, using the same phase shift parameterization, it will be interesting to analyze the mass and width of Δ(1232) resonance, which are commonly known as Breit-Wigner mass and Breit-Wigner width, respectively.